\documentclass[conference]{IEEEtran}
\IEEEoverridecommandlockouts
\PassOptionsToPackage{table,xcdraw}{xcolor}
\usepackage{cite}
\usepackage{pgf}
\usepackage{collcell}
\usepackage{amsmath,amssymb,amsfonts}
\usepackage{algorithmic}
\usepackage{graphicx}
\usepackage[warn]{textcomp}
\usepackage{xcolor}
\usepackage{multirow}
\usepackage{tikz}
\def\BibTeX{{\rm B\kern-.05em{\sc i\kern-.025em b}\kern-.08em
    T\kern-.1667em\lower.7ex\hbox{E}\kern-.125emX}}

\usepackage{verbatim}
\usepackage[font={small}]{caption}
\usepackage{subcaption}
\usepackage{hyperref}
\usepackage[]{algorithm2e}
\usepackage{amsmath}
\usepackage{multirow, tabularx}
\newcommand{\comments}[1]{}
\begin{document}

\title{The Performance Evaluation of Attention-Based Neural ASR under Mixed Speech Input }

\author{
\IEEEauthorblockN{1\textsuperscript{st} Bradley He}
\IEEEauthorblockA{\textit{Dept. Computer Science} \\
\textit{Stony Brook University, NY, USA}\\
bhe1@student.gn.k12.ny.us} \\
\and
\IEEEauthorblockN{2\textsuperscript{nd} Martin Radfar}
\IEEEauthorblockA{\textit{Dept. Computer Science} \\
\textit{Stony Brook University, NY, USA}\\
radfar@cs.stonybrook.edu}
}

\maketitle

\begin{abstract}
 In order to evaluate the performance of the attention based neural ASR under noisy conditions, the current trend is to present hours of various noisy speech data to the model and measure the overall word/phoneme error rate (W/PER).  In general, it is unclear how these models perform when exposed to a cocktail party setup in which two or more speakers are active.  In this paper, we present the mixtures of speech signals to a popular attention-based neural ASR, known as Listen, Attend, and Spell (LAS), at different target-to-interference ratio (TIR) and measure the phoneme error rate.  In particular, we investigate in details when two phonemes are mixed what will be the predicted phoneme; in this fashion we build a model in which the most probable predictions for a phoneme are given. We found a 65\% relative increase in PER when LAS was presented with mixed speech signals at TIR = 0 dB and the performance approaches the unmixed scenario at TIR = 30 dB.  Our results show the model, when presented with mixed phonemes signals, tend to predict those that have higher accuracies during evaluation of original phoneme signals. 
 
\end{abstract}


\section{Introduction}
The neural ASR technology cannot be ubiquitously used as a reliable human-machine interface unless the users trust the technology in highly non-stationary noisy environments \cite{ zhang2018deep,barker2013pascal}. The multi-speaker environment, AKA the cocktail party problem\cite{haykin2005cocktail}, is one of the adverse cases for the usage of ASR technology.  When dealing with neural ASR performance assessment under the multi-speaker scenario, there are two groups of works \footnote{Here we only consider a single channel recording and one microphone at the scene}: The first group treats this problem as  source separation by inferring the target source from the mixture and assessing the ASR output after the separation task \cite{ huang2015joint,du2016regression}.  The separation process is either pre-, or post-integrated into neural ASR model. Due to complexity and lack of generalization, these models have not been used in any ASR technologies so far to the best of our knowledge. The second group trains the neural ASR with a large corpus of noisy speech data. These models tend to learn all noisy patterns associated with a label and usually deliver good performance, are less complex and noise-independent, and are commonly used in today’s robust ASR technologies \cite{chiu2018state}.  While these models have been tested on tens of hours of noisy data, no study has been contacted to specifically assess their performance with mixed input speech signals in different target-to interference ratio to the best of our knowledge.  In this paper, we investigate how attention based neural ASR models \cite{chan2015las, garofolo1992timit}---we adapt the model proposed in \cite{chan2015las }, known as Listen, Attend, and Spell (LAS)---performs when exposed to mixed speech signals. We mixed speech signals with different target-to-interference ratio (TIR) from 0 dB to 30 dB in 3 dB increments, present them to the ASR model, and predict the output phonemes.  We also segment test speech signals into phonemes, mix various combinations of phonemes, and measure the performance of the ASR model on the mixed phoneme signals.

We find in our speech mixing experiments that the intensity of interference noise impacts the performance of LAS on a mixed speech signal, with lower TIRs producing high PER values and high TIRs producing near-optimal PER values. We also observe that LAS performs slightly worse on voice mixtures with a female target as compared to voice mixtures with a male target. This suggests that disproportionate gender ratios in the training phase are picked up by LAS and carried over during evaluation. Our phoneme mixing experiments found that phonemes which are accurately predicted by LAS also tend to perform well in a mixed context, suggesting that accuracy rates of unmodified phoneme signals are an indicator of the phoneme's ability to be predicted in a mixed context.

The rest of this paper organized as follows: The basic LAS architecture and various modifications made to it are described in Section II. Our experiments, methodology, and findings on mixed speech signals are described in Section III and our experiments, methodology, and findings on mixed phoneme signals are described in Section IV. A final summary of our work is given in Conclusions.

\section{Model Description}

\subsection{Listen, Attend, and Spell}
The model that was used throughout this experiment is the recently developed Listen, Attend, and Spell (LAS) presented in \cite{zhang2018deep,chan2015las}, a neural network that takes in acoustic features as an input and outputs English characters. The underlying framework of LAS is composed of two RNNs: a pyramidal encoder RNN,  known as the listener, and a decoder RNN, known as the speller. The listener module maps the low-level features in the input to high-level features. Between the listener and speller step is the attender, which examines the high-level features produced by the listener module to generate a context vector. The speller module takes the high-level features produced by the listener module and the context vector produced by the attender to produce a character probability distribution.

\subsection{Modifications}
Some differences are present between the model used in the experiment and the model presented in the original paper.  Label smoothing was implemented in the same manner as in \cite{chiu2017label}. Label smoothing encourages the model to be less confident in its predictions by increasing the entropy of the model during prediction. This reduces overfitting during training and validation and allows the model to be more adaptable \cite{szegedy2015label}. Modifications also had to be made due to the nature of the TIMIT data set. Instead of predicting character sequences, the model was configured to interpret and predict phoneme sequences, which has been successfully done with  other ASR models \cite{yeh2018unsupervised}. The LAS model with these changes implemented was provided to us by Alexander H. Liu, whose code is available online\footnote{\url{https://github.com/AzizCode92/Listen-Attend-and-Spell-Pytorch}}.

\begin{figure}
	\includegraphics[width=\columnwidth]{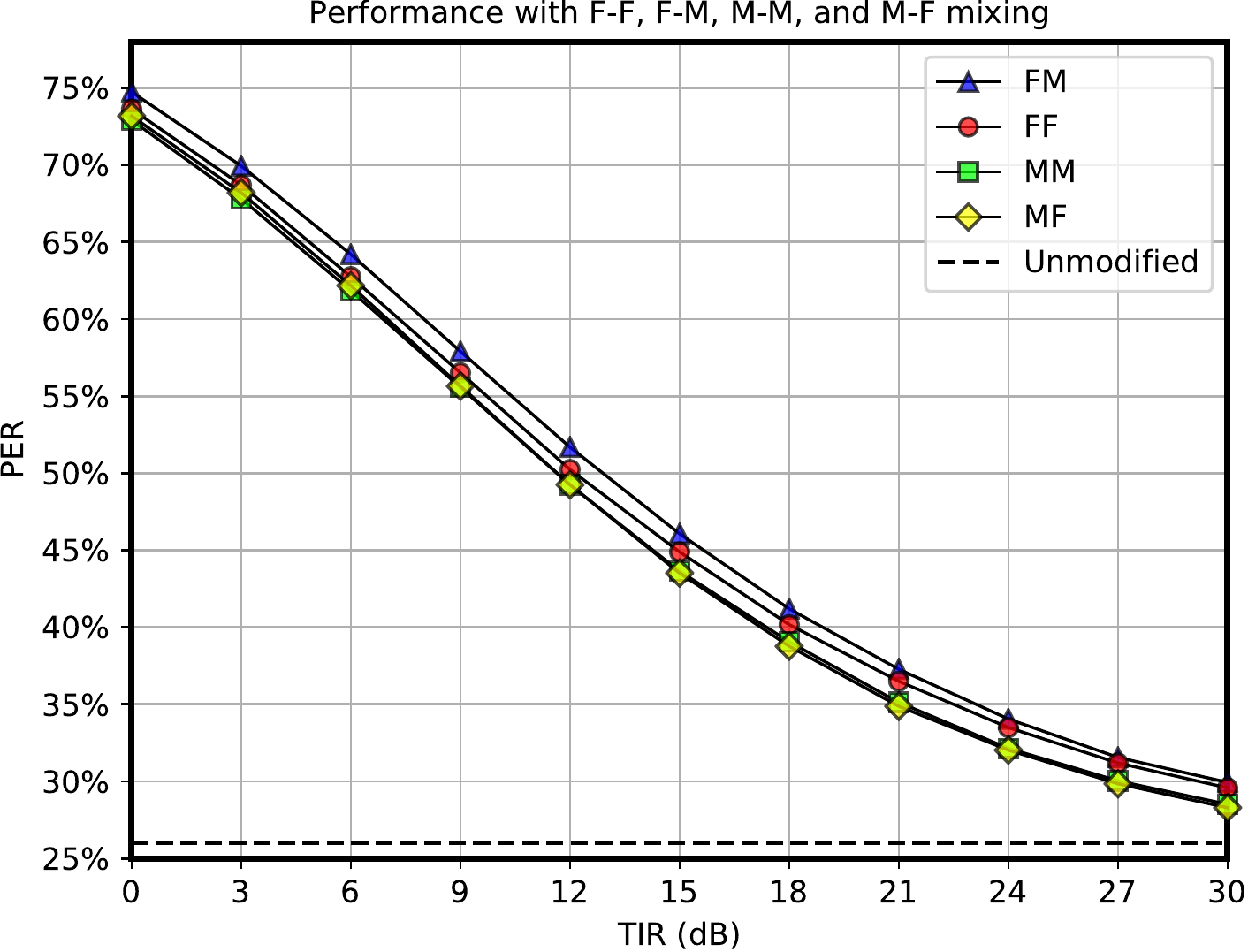}
	\caption{ PER versus TIR of female-male mixing ($\triangle$ line), female-female mixing ($\circ$ line), male-male mixing ($\square$ line), and male-female mixing ($\diamond$ line) averaged over 33 sets compared with PER of unmodified testing set (dashed line).}
	\label{all}
\end{figure}

\section{Voice Mixing Experiment}

\subsection{Experimental Setup}
Speech samples utilized in the experiment were obtained from the TIMIT data set. The training data set consisted of 462 speakers, 326 of whom were male and 136 of whom were female, where each speaker uttered 10 sentences. For training, validation, and testing, the speech samples were converted from WAVE format to a vector of 39 coefficients consisting of its MFCCs plus energy, their first derivative, and their second derivative, as described in \cite{graves2012preprocess}. For evaluation, the phoneme sequence outputted by the model and the ground truth sequence were collapsed from 61 into 39 phonemes \cite{lopes2011phoneme}.

 The testing data set consisted of 168 speakers, 112 of whom were male and 56 of whom were female, each of whom uttered 10 sentences. Speech samples were mixed in four target-interference combinations: male-male, female-female, male-female, and female-male. Each gender combination was mixed at TIRs from 0 dB to 30 dB in 3 dB increments. A testing set for a specific gender combination and TIR was generated by taking all sentences uttered by the target gender that appeared in the testing data set and mixing them with a randomly-chosen sentence of the appropriate gender. 33 of such testing sets were generated for each gender combination and TIR\footnote{Code is available at \url{https://github.com/BradleyHe/TIMIT-Voice-Mixer}}.

\subsection{Results}
	To evaluate the performance of the model, the phoneme error rate (PER), defined as the word error rate of the predicted phoneme sequence as compared to the base truth phoneme sequence, was calculated for all mixed voice signals evaluated by the model. Averaged PER results are presented in Fig. \ref{all} and Table \ref{table:tirtable}. Fig. \ref{all} compares PER against TIR for female-female, female-male, male-male, and male-female mixtures. Additionally, the performance of the optimal trained model on the unmodified testing data set, a 26\% phoneme error rate, is represented on Fig. \ref{all} by the dashed lines. Table \ref{table:tirtable} displays the average PER of the four gender combinations at TIRs of 0, 15, and 30 db.
	
\newcolumntype{Y}{>{\centering\arraybackslash}X}
\setlength\extrarowheight{2pt}
\begin{table}[h]
	\centering
	\caption{Phoneme error rates of the 4 possible target-interference gender combinations at selected TIRs of 0, 15, and 30 db.}
	\label{table:tirtable}
	\begin{tabularx}{.8\columnwidth}{|
	 >{\hsize=1.3\hsize}Y| 
	 >{\hsize=0.9\hsize}Y|
	 >{\hsize=0.9\hsize}Y|
	 >{\hsize=0.9\hsize}Y|}
	 
	\hline
	\multirow{2}{*}{Gender Mixing} 
	&\multicolumn{3}{c|}{Target-Interference Ratio}\\
	\cline{2-4}
	&0 db      &15 db 	&30 db\\
	\hline
	F-M 	 &74.74\%   &46.07\%   &29.94\%\\
	\hline
	F-F      &73.63\%   &44.89\%   &29.58\%\\
	\hline
	M-M      &72.91\%   &43.64\%   &28.51\%\\
	\hline
	M-F      &73.18\%   &43.52\%   &28.29\%\\
	\hline
	\end{tabularx}
\end{table}	
	
\setlength\tabcolsep{4pt}
\setlength\extrarowheight{2pt}
\begin{table}[t]
\centering
\caption{Prediction rates on phoneme mixtures in the complete phoneme set. Phonemes are sorted by stratification accuracy. Values in cells are the prediction rate of the phoneme indicated by the row when mixed with the phoneme indicated by the column. }

\label{table:completeaccuracy}
\begin{tabular}{|l|l|l|l|l|l|l|l|l|l|l|}
\hline
   & ow                           & ey                           & ah                           & ay                           & er                           & s                            & t                            & aa                           & ih                           & eh                           \\ \hline
ow & \cellcolor[HTML]{999999}75.6 & \cellcolor[HTML]{DDDDDD}32.5 & \cellcolor[HTML]{BBBBBB}56.7 & \cellcolor[HTML]{DDDDDD}22.8 & \cellcolor[HTML]{BBBBBB}48.3 & 10.0                         & \cellcolor[HTML]{BBBBBB}57.4 & \cellcolor[HTML]{BBBBBB}52.0 & \cellcolor[HTML]{999999}60.3 & \cellcolor[HTML]{BBBBBB}45.3 \\ \hline
ey & \cellcolor[HTML]{DDDDDD}32.4 & \cellcolor[HTML]{999999}77.6 & \cellcolor[HTML]{BBBBBB}46.9 & \cellcolor[HTML]{DDDDDD}24.7 & \cellcolor[HTML]{DDDDDD}32.6 & 12.6                         & \cellcolor[HTML]{DDDDDD}30.4 & 16.4                         & \cellcolor[HTML]{BBBBBB}49.3 & \cellcolor[HTML]{BBBBBB}40.4 \\ \hline
ah & \cellcolor[HTML]{DDDDDD}34.9 & \cellcolor[HTML]{DDDDDD}21.6 & \cellcolor[HTML]{999999}65.7 & \cellcolor[HTML]{BBBBBB}44.6 & \cellcolor[HTML]{BBBBBB}47.3 & \cellcolor[HTML]{DDDDDD}35.3 & \cellcolor[HTML]{DDDDDD}28.5 & \cellcolor[HTML]{BBBBBB}47.9 & \cellcolor[HTML]{BBBBBB}41.4 & \cellcolor[HTML]{999999}65.3 \\ \hline
ay & \cellcolor[HTML]{DDDDDD}32.0 & \cellcolor[HTML]{DDDDDD}28.6 & \cellcolor[HTML]{BBBBBB}42.3 & \cellcolor[HTML]{BBBBBB}58.9 & \cellcolor[HTML]{DDDDDD}28.8 & 10.1                         & \cellcolor[HTML]{DDDDDD}28.8 & \cellcolor[HTML]{DDDDDD}38.7 & \cellcolor[HTML]{DDDDDD}37.7 & \cellcolor[HTML]{DDDDDD}30.2 \\ \hline
er & 13.8                         & 4.5                          & \cellcolor[HTML]{DDDDDD}29.0 & 11.7                         & \cellcolor[HTML]{BBBBBB}55.1 & 7.7                          & \cellcolor[HTML]{BBBBBB}41.5 & 16.3                         & \cellcolor[HTML]{DDDDDD}29.6 & 16.9                         \\ \hline
s  & 19.2                         & 5.1                          & \cellcolor[HTML]{DDDDDD}22.7 & 11.4                         & 16.2                         & \cellcolor[HTML]{999999}60.7 & \cellcolor[HTML]{DDDDDD}37.0 & \cellcolor[HTML]{DDDDDD}20.2 & \cellcolor[HTML]{DDDDDD}26.1 & 16.8                         \\ \hline
t  & \cellcolor[HTML]{DDDDDD}28.3 & 13.5                         & 8.7                          & 19.4                         & 16.8                         & \cellcolor[HTML]{BBBBBB}56.9 & \cellcolor[HTML]{BBBBBB}47.0 & \cellcolor[HTML]{DDDDDD}34.1 & 6.0                          & \cellcolor[HTML]{DDDDDD}20.7 \\ \hline
aa & 7.5                          & 2.5                          & 11.3                         & 6.0                          & 8.4                          & 3.7                          & 10.2                         & 17.4                         & 11.2                         & 6.0                          \\ \hline
ih & 2.2                          & 14.1                         & 6.9                          & 1.5                          & 15.6                         & \cellcolor[HTML]{BBBBBB}46.9 & 16.4                         & 2.1                          & \cellcolor[HTML]{DDDDDD}28.0 & 8.8                          \\ \hline
eh & 8.6                          & 17.7                         & 5.5                          & 2.4                          & 7.7                          & 2.4                          & 5.8                          & 5.8                          & 15.2                         & 15.4                         \\ \hline
\end{tabular}
\end{table}	

	The first observation is that the average PER declines as the TIR increases, which is clear in Fig. \ref{all}. All four gender combinations have a PER of around 72.5\% at a TIR of 0 db, which steadily decreases to around 29\% at a TIR of 30 db. This is expected, as the interference voice becomes less intense and therefore interferes less with the target voice, which allows the phonemes of the target voice to be predicted more accurately. We can also see that the improvement in PER within a given 3 db interval becomes smaller and smaller as the PER approaches the unmodified testing data set performance of 26\%. This is also expected, as the difference in intensity of the interference voice eventually becomes negligible at higher TIR values.

	Although the average PERs of the four gender mixtures keep relatively close across all TIRs, we still notice a difference between the mixtures. The model performs around 1\% PER worse on mixed speech samples with female targets than on those with male targets. This disparity is also present in the original testing set; the model has a 26.5\% PER on the unmodified testing set with all male speakers removed, as compared to the 26\% PER on the complete testing set. This can be attributed to the training stage of the model; the training data set has a disproportionate ratio of male to female speakers, with more than twice the amount of male speakers as compared to female speakers. Because of this, the model might have developed a stronger ability to interpret male voices as compared to female voices.

\section{Phoneme Mixing Experiment}

\subsection{Experimental Setup}
A total of 51368 non-silence phonemes were extracted from the complete TIMIT testing data set. These phonemes were collapsed from 63 phonemes into 39 phonemes as described in \cite{lopes2011phoneme}. Table \ref{table:phntable1} shows the 10 most frequent phonemes in the complete phoneme set. The LAS model described in the previous experiment was used to evaluate each phoneme signal and remove phoneme signals that were not correctly identified by the model, which generated a "stratified phoneme set". This was done by evaluating a given phoneme signal and checking if the prediction included the correct phoneme in it. Table \ref{table:phntable2} shows the percentage of signals correctly predicted by the model during stratification, or "stratification accuracy", of the 10 most frequent phonemes, as well as their occurrences.

The 10 phonemes listed in table \ref{table:phntable2}, which also have the highest stratification accuracies, were tested on in this experiment. Phoneme mixing was performed by randomly selecting two phoneme samples from the specified phoneme set, mixing them at a TIR of 0 db, and evaluating the mixed signal using the LAS model. 2000 phoneme mixings were performed for each unique combination of phonemes in both the stratified phoneme set and the complete phoneme set\footnote{Code is available at \url{https://github.com/BradleyHe/TIMIT-Phoneme-Mixer}}.

\setlength\tabcolsep{4.0pt}
\begin{table}[]
\centering
\caption{Prediction rates on phoneme mixtures in the stratified phoneme set. Phonemes are sorted by stratification accuracy. Values in cells are the prediction rate of the phoneme indicated by the row when mixed with the phoneme indicated by the column. }
\label{table:stratifiedaccuracy}
\begin{tabular}{|l|l|l|l|l|l|l|l|l|l|l|}
\hline
   & ow                           & ey                           & ah                           & ay                           & er                           & s                            & t                            & aa                           & ih                           & eh                           \\ \hline
ow & \cellcolor[HTML]{777777}91.7 & \cellcolor[HTML]{DDDDDD}38.4 & \cellcolor[HTML]{999999}63.7 & 18.8                         & \cellcolor[HTML]{BBBBBB}46.6 & 7.9                          & \cellcolor[HTML]{BBBBBB}57.2 & \cellcolor[HTML]{999999}61.6 & \cellcolor[HTML]{BBBBBB}59.5 & \cellcolor[HTML]{BBBBBB}44.5 \\ \hline
ey & \cellcolor[HTML]{BBBBBB}40.2 & \cellcolor[HTML]{777777}96.0 & \cellcolor[HTML]{BBBBBB}56.0 & \cellcolor[HTML]{DDDDDD}24.9 & \cellcolor[HTML]{DDDDDD}29.0 & 18.4                         & \cellcolor[HTML]{DDDDDD}34.9 & 6.6                          & \cellcolor[HTML]{999999}61.4 & \cellcolor[HTML]{BBBBBB}55.6 \\ \hline
ah & \cellcolor[HTML]{DDDDDD}34.5 & \cellcolor[HTML]{DDDDDD}26.8 & \cellcolor[HTML]{777777}93.6 & \cellcolor[HTML]{DDDDDD}31.1 & \cellcolor[HTML]{999999}60.2 & \cellcolor[HTML]{999999}63.0 & \cellcolor[HTML]{DDDDDD}39.0 & 18.8                         & \cellcolor[HTML]{BBBBBB}46.5 & \cellcolor[HTML]{BBBBBB}59.1 \\ \hline
ay & \cellcolor[HTML]{BBBBBB}49.8 & \cellcolor[HTML]{DDDDDD}36.4 & \cellcolor[HTML]{999999}66.2 & \cellcolor[HTML]{777777}87.3 & \cellcolor[HTML]{DDDDDD}38.2 & 9.0                          & \cellcolor[HTML]{DDDDDD}35.2 & \cellcolor[HTML]{BBBBBB}49.7 & \cellcolor[HTML]{BBBBBB}44.9 & \cellcolor[HTML]{DDDDDD}39.4 \\ \hline
er & \cellcolor[HTML]{DDDDDD}23.3 & 5.5                          & \cellcolor[HTML]{DDDDDD}33.7 & 14.5                         & \cellcolor[HTML]{777777}86.7 & 3.4                          & \cellcolor[HTML]{BBBBBB}53.0 & 19.3                         & \cellcolor[HTML]{DDDDDD}25.4 & 12.9                         \\ \hline
s  & \cellcolor[HTML]{DDDDDD}34.2 & 6.1                          & \cellcolor[HTML]{BBBBBB}47.0 & \cellcolor[HTML]{DDDDDD}20.2 & \cellcolor[HTML]{DDDDDD}29.4 & \cellcolor[HTML]{777777}98.4 & \cellcolor[HTML]{999999}68.7 & \cellcolor[HTML]{DDDDDD}38.0 & \cellcolor[HTML]{DDDDDD}35.1 & \cellcolor[HTML]{DDDDDD}26.4 \\ \hline
t  & \cellcolor[HTML]{BBBBBB}51.6 & \cellcolor[HTML]{DDDDDD}27.2 & \cellcolor[HTML]{DDDDDD}21.3 & \cellcolor[HTML]{DDDDDD}34.0 & \cellcolor[HTML]{DDDDDD}38.4 & \cellcolor[HTML]{999999}68.2 & \cellcolor[HTML]{777777}89.7 & \cellcolor[HTML]{BBBBBB}54.3 & 8.5                          & \cellcolor[HTML]{BBBBBB}46.0 \\ \hline
aa & \cellcolor[HTML]{DDDDDD}26.1 & 9.9                          & \cellcolor[HTML]{BBBBBB}47.0 & 15.7                         & \cellcolor[HTML]{DDDDDD}28.0 & 8.1                          & \cellcolor[HTML]{BBBBBB}45.0 & \cellcolor[HTML]{999999}76.7 & \cellcolor[HTML]{DDDDDD}36.3 & 19.9                         \\ \hline
ih & 9.2                          & 19.8                         & 18.0                         & 1.6                          & \cellcolor[HTML]{DDDDDD}38.4 & \cellcolor[HTML]{777777}96.0 & \cellcolor[HTML]{BBBBBB}57.3 & 4.3                          & \cellcolor[HTML]{777777}91.9 & \cellcolor[HTML]{BBBBBB}55.4 \\ \hline
eh & \cellcolor[HTML]{DDDDDD}30.2 & \cellcolor[HTML]{DDDDDD}33.5 & \cellcolor[HTML]{DDDDDD}23.7 & 3.5                          & \cellcolor[HTML]{DDDDDD}23.6 & 12.1                         & \cellcolor[HTML]{DDDDDD}25.4 & 11.0                         & \cellcolor[HTML]{DDDDDD}29.8 & \cellcolor[HTML]{777777}85.4 \\ \hline
\end{tabular}
\end{table}

\setlength\tabcolsep{4.0pt}
\begin{table}[!htb]
	\centering
    \begin{minipage}{.40\columnwidth}
      \caption{Occurrences of the 10 most frequent phonemes in the complete phoneme set.}
      \label{table:phntable1}
      \centering
        \begin{tabular}{|c|c|} 
		\hline
		Phoneme & Occurrences \\
		\hline
		ih & 4654 \\ 
		n & 3112 \\ 
		iy & 2710 \\ 
		l & 2699 \\
		s & 2639 \\
		r & 2525 \\ 
		ah & 2343 \\ 
		aa & 2289 \\ 
		er & 2183 \\ 
		k & 1614 \\
		\hline
		
		\end{tabular}
		
    \end{minipage}\hfill
    \begin{minipage}{.54\linewidth}
    \caption{Occurrences and stratification accuracy of the 10 most frequent phonemes in the stratified phoneme set.}
    \label{table:phntable2}
      \centering
		\begin{tabular}{|c|c|c|}
		\hline
		Phoneme & Occurrences & Accuracy \\
		\hline
		ah & 1278 & 54.5\% \\ 
		s & 1115 & 42.3\%\\ 
		er & 1115 & 51.1\% \\ 
		ih & 840 & 18.0\% \\
		t & 620 & 40.4\% \\
		ow & 545 & 70.1\% \\ 
		ey & 514 & 63.8\% \\ 
		ay & 464 & 54.5\% \\ 
		aa & 414 & 18.1\% \\ 
		eh & 202 & 14.0\% \\
		\hline
		\end{tabular}
    \end{minipage} 
\end{table}

\subsection{Results}
\begin{figure}[]
	\includegraphics[width=\columnwidth]{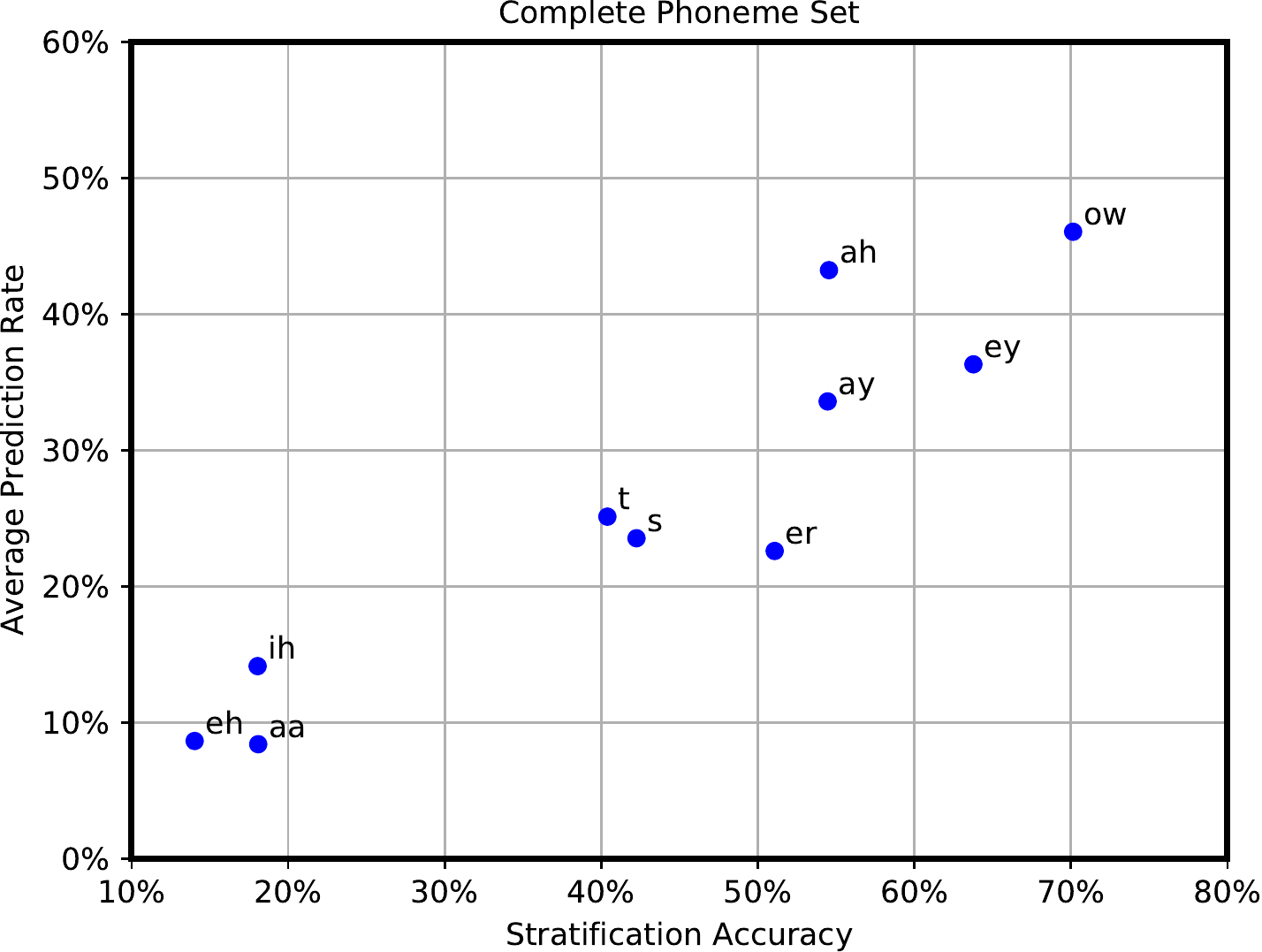}
	\caption{ Scatter plot of the 10 phonemes tested in this experiment. X-axis indicates stratification accuracy, whereas the y-axis indicates the average prediction rate across all mixtures of a phoneme in the complete phoneme set. }
	\label{scattercomplete}
\end{figure}

\begin{figure}[]
	\includegraphics[width=\columnwidth]{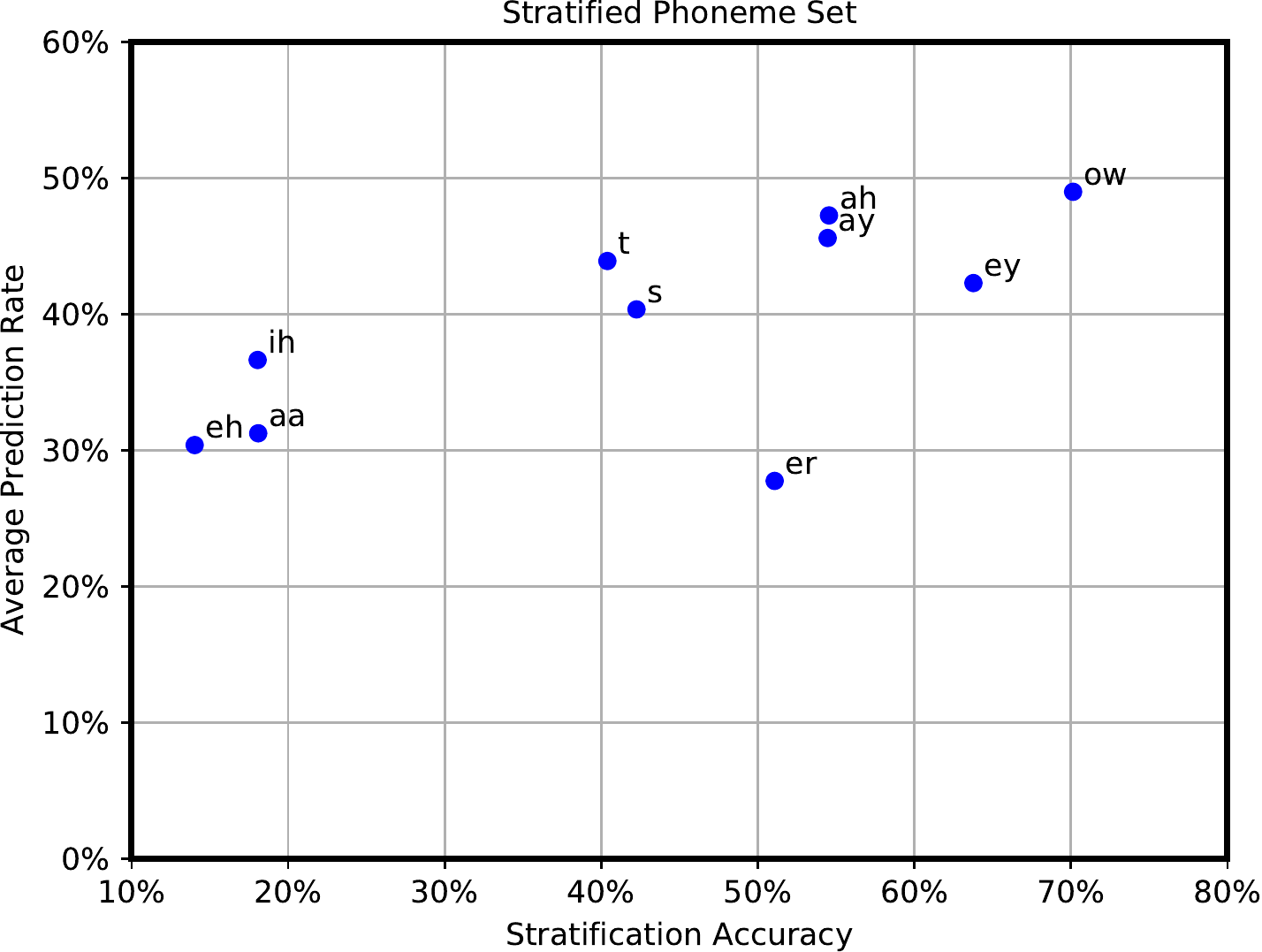}
	\caption{ Scatter plot of the 10 phonemes tested in this experiment. X-axis indicates stratification accuracy, whereas the y-axis indicates the average prediction rate across all mixtures of a phoneme in the stratified phoneme set. }
	\label{scatterstratified}
\end{figure}

Multiple metrics were used to determine the performance of the LAS model on the mixed phoneme signals. The "prediction rate" of a target phoneme when mixed with another phoneme is determined by finding the percentage of predictions generated by the LAS model that contain the target phoneme. Tables \ref{table:completeaccuracy} and \ref{table:stratifiedaccuracy} depict the prediction rate of the 10 phonemes in every possible combination in the complete phoneme set and the stratified phoneme set, respectively. The "error rate" of a mixture of phonemes is determined by finding the percentage of predictions that do not contain either of the two phonemes mixed together. A pair of phonemes is said to be "accuracy oriented" if the phoneme with the higher stratification accuracy out of the two also had a higher prediction rate than the other when mixed. If a pair of phonemes was accuracy oriented, it meant that the ability of the model to recognize a certain phoneme over the noise of another phoneme was linked to the stratification accuracies of the two phonemes. Table \ref{table:statistics} contains the average error rate of all combinations, the average prediction length, and the number of phoneme combinations that were accuracy aligned in the two phoneme sets. Finally, Figs. \ref{scattercomplete} and \ref{scatterstratified} compare the phonemes' average prediction rate across all of their mixtures against their stratification accuracy. 

Table \ref{table:statistics} reveals that the stratification process improved the model's ability to recognize mixed phoneme signals, bringing the average error rate down by around 20\%. The average prediction length increased, which was likely caused by the stratification process removing phoneme signals from which the model generated empty predictions. The increase in performance from the complete set to the stratified set can also be seen in Tables \ref{table:completeaccuracy} and \ref{table:stratifiedaccuracy}, where the stratified set has generally higher prediction rates than the complete set. This is expected, as removing phoneme signals that the model does not correctly predict will ensure that mixed signals will be comprised of phoneme signals that the model has already correctly predicted. The dark diagonal stripes seen in Tables \ref{table:completeaccuracy} and \ref{table:stratifiedaccuracy} showcase the high prediction rates of phoneme mixtures consisting of the same phoneme, which suggests that mixing two similar phonemes results in an audio signal that is more comprehensible to the LAS model.

\setlength\tabcolsep{4.0pt}
\begin{table}[!htb]
\centering
\caption{Selected statistics from testing the two phoneme sets. }
\label{table:statistics}
\begin{tabular}{|c|c|c|c|}
\hline
Set & Avg. Length & Avg. Error Rate & Accuracy Oriented \\
\hline
Complete & 1.10 phns & 53.7\% & 36/45 (80\%) \\
\hline
Stratified & 1.27 phns & 32.4\% & 30/45 (66.7\%)\\
\hline
\end{tabular}
\end{table}

Table \ref{table:statistics} also showcases the relationship between stratification accuracy and prediction rate. In the complete set, this relationship is most prominent, with 36/45 (80\%) of non-duplicate phoneme mixtures being accuracy oriented. This is expected, since the stratification accuracy directly impacts the percentage of phonemes that the model is able to correctly predict in the complete set. However, the stratified set exhibits this behavior as well. 30/45 (66.7\%) of non-duplicate phoneme mixtures in the stratified set are accuracy oriented. Although the percentage of accuracy oriented mixtures has decreased, the majority of mixtures still exhibits this behavior.

This relationship is more clearly seen in Figs. \ref{scattercomplete} and \ref{scatterstratified}, where the average prediction rate for each phoneme is plotted against its stratification accuracy. In Fig. \ref{scattercomplete}, a clear relationship is seen between the stratification accuracy of a phoneme and its prediction rate within the complete set. In the stratified set, Fig. \ref{scatterstratified} still exhibits a direct relationship, albeit with a smaller slope. Although this relationship can be explained in the complete set, since phoneme signals that cannot be identified by the model are still present, the same cannot be said about the stratified set, where those phoneme signals have been removed. The relationship between stratification accuracy and prediction rate within the stratified set suggests that the ability of the LAS model to identify certain phonemes is linked to how well the LAS model can identify that phoneme over interference noise. It also suggests that some phonemes are inherently more recognizable than others and will consequently be identified over other phonemes in a mixed environment, more often than not.

\section{Conclusions}
In this paper, we test the performance of Listen, Attend, and Spell on various mixed voice and phoneme signals from the TIMIT data set. Testing on mixed voice signals at different TIRs reveals that the intensity of interference noise directly impacts the performance of LAS, with higher TIRs resulting in higher PERs. Additionally, gender proportions during the training phase was found to impact performance, where the more prevalent gender had lower PERs than the less prevalent gender. Testing on mixed phoneme signals reveals that the model is able to accurately predict phonemes that have higher stratification accuracies, more so than those with lower stratification accuracies, in a mixed signal. This suggests that the performance of LAS on a phoneme's unmodified signal indicates LAS's familiarity with that phoneme, which translates into better performance when predicting mixed signals.

\bibliographystyle{IEEEbib}
\bibliography{refs1}

\end{document}